\newif\ifpdf
\ifx\pdfoutput\undefined
\pdffalse 
\else
\pdfoutput=1 
\pdftrue \fi

\documentclass[figure]{epl}
\usepackage{amsmath}
\usepackage{amssymb}
\usepackage{amsthm}
\usepackage{amsfonts}
\usepackage{verbatim}
\usepackage{mathrsfs}

\DeclareFontFamily{OT1}{eusb}{} \DeclareFontShape{OT1}{eusb}{m}{n}
{<5> <6> <7> <8> <9> <10> <11> <12> <14.4> eusb10}{}
\DeclareMathAlphabet{\eusb}{OT1}{eusb}{m}{n}

\DeclareFontFamily{OT1}{eusm}{} \DeclareFontShape{OT1}{eusm}{m}{n}
{<5> <6> <7> <8> <9> <10> <11> <12> <14.4> eusm10}{}
\DeclareMathAlphabet{\eusm}{OT1}{eusm}{m}{n}

\DeclareFontFamily{OT1}{eufm}{} \DeclareFontShape{OT1}{eufm}{m}{n}
{<5> <6> <7> <8> <9> <10> <11> <12> <14.4> eufm10}{}
\DeclareMathAlphabet{\mathfrak}{OT1}{eufm}{m}{n}

\newcommand{\textd}{\text{\rm d}}

\newcommand{\BbbP}{\mathbb{P}}

\renewcommand{\AA}{\mathcal{A}}
\newcommand{\BB}{\mathcal{B}}
\newcommand{\GG}{\mathcal{G}}

\newcommand{\bS}{\boldsymbol S}
\newcommand{\bs}{\boldsymbol s}
\newcommand{\bk}{\boldsymbol k}
\newcommand{\br}{\boldsymbol r}

\newcommand{\hata}{\hat{\text{\rm a}}}
\newcommand{\hatb}{\mkern-4mu\hat{\mkern4mu\text{\rm b}}}
\newcommand{\hatc}{\hat{\text{\rm c}}}
 \newcommand{\hate}{\hat{\text{\rm e}}}

\newcommand{\zz}{\mathfrak{z}}

\newcommand{\tg}{\textsl{t}_{\textsl{2g}}}
\newcommand{\eg}{\textsl{e}_{\textsl{g}}}
\newcommand{\td}{\textsl{3d}}
\newcommand{\tp}{\textsl{4p}}

\newcommand{\SW}{\mathchoice%
{\text{\small\rm SW}}
{\text{\small\rm SW}}
{\text{\tiny\rm SW}}
{\text{\tiny\rm SW}}
}

\newcommand{\BBE}{\BB_{\text{\rm E}}}
\newcommand{\BBSW}{\BB_{\text{\rm SW}}}

\newcommand{\twoeqref}[2]{Eqs.~(\ref{#1}-\ref{#2})}

\title{\Large Orbital order in classical models of transition-metal\\compounds}
\shorttitle{Orbital order in TM compounds}
\author{Z.~Nussinov\inst{1} \and M.~Biskup\inst{2} \and L.~Chayes\inst{2} \and  J.~van den Brink\inst{3}}
\shortauthor{Nussinov \etal}
\institute{
  \inst{1} Theoretical Division, Los Alamos National Laboratory, Los Alamos, NM 87545, USA\\ 
  \inst{2} Department of Mathematics, UCLA, Los Angeles CA 90095-1555, USA\\ 
  \inst{3} Lorentz Institute for Theoretical Physics, Universiteit Leiden, Postbus 9506, NL-2300 RA Leiden, The Netherlands
}

\begin{document}

\maketitle
\begin{abstract}
We study the classical 120-degree and related orbital models. These are the classical limits of quantum models which describe the interactions among orbitals of transition-metal compounds. We demonstrate that at low temperatures these models exhibit a long-range order which arises via an ``order by disorder'' mechanism. This strongly indicates that there is orbital ordering in the quantum version of these models, notwithstanding recent rigorous results on the absence of spin order~in~\hbox{these systems}.
\end{abstract}

\section{Introduction}
The properties of transition-metal (TM) compounds are a topic of long-standing interest. In these materials, the fractional filling of the $\td$-shells in the TM ion provides a novel facet: The splitting of the $\tg$ and $\eg$ orbitals by the crystal field can produce situations with a single dynamical electron (or a hole) on each site along with multiple orbital degrees of freedom \cite{Goodenough-book,Kugel73,Imada98}. Pertinent examples are found among the vanadates (e.g., V$_2$O$_3$~\cite{Castellani78}, LiVO$_2$~\cite{Pen97}, LaVO$_3$~\cite{Khaliullin01}), cuprates (e.g., KCuF$_3$~\cite{Kugel73}) and derivatives of the colossal magnetoresistive manganite LaMnO$_3$~\cite{brink03}. The presence of the extra degrees of freedom raises the theoretical possibility of global, cooperative effects; i.e., \emph{orbital ordering}. Such ordering may be observed via associated orbital-related magnetism and lattice distortions or, e.g., by resonant X-ray scattering techniques in which the~$\td$ orbital order is detected by its effect on excited~$\tp$ states~\cite{RXRays}. 

The case for orbital ordering has been bolstered by detailed calculations and various other considerations~\cite{OrbitalOrder}.
However, alternate perspectives and various conceptual doubts have been raised concerning the entire picture of long-range orbital ordering~\cite{Feiner97,Harris03}. In particular, at the theoretical level, a satisfactory justification of orbital ordering has not yet been provided~\cite{casimir}.
The goal of this Letter is to present arguments which irrefutably demonstrate that orbital ordering indeed occurs. We will discuss primarily the so-called \emph{120$^\circ$-model} which
describes the situations when the~$\eg$ orbitals are occupied by a single electron.

In the TM-compounds, the initial point of all derivations is to \emph{neglect} the strain-field induced interactions among orbitals. Starting from the appropriate itinerant electron model, a standard super-exchange calculation leads to the Kugel-Khomskii model \cite{Kugel73} with the Hamiltonian given by
\begin{equation}
\label{1}
H = \sum_{\langle\br,\br' \rangle} H_{\text{orb}}^{\br,\br'} 
\bigl(\bs_{\br}\cdot \bs_{\br'} + \tfrac14\bigr).
\end{equation}
Here $\bs_{\br}$ denotes the spin of the electron at site~$\br$ and $H_{\text{orb}}^{\br,\br'}$ are operators acting on the orbital degrees of freedom.
For the TM-atoms arranged in a cubic lattice, these take the form 
\begin{eqnarray}
\label{orbHam}
H_{\text{orb}}^{\br,\br'} = J(4\hat\pi_{\br}^\alpha \hat\pi_{\br'}^\alpha 
-2\hat\pi_{\br}^\alpha - 2\hat\pi_{\br'}^\alpha+1),
\end{eqnarray}
where the~$\hat\pi_{\br}^\alpha$ denote orbital pseudospin operators acting on the appropriate orbital multiplet and~$\alpha=x,y,z$ is the direction of the bond~$\langle\br,\br'\rangle$. 
In the~$\eg$ compounds, we have
\begin{equation}
\hat\pi_{\br}^x=\tfrac14(-\sigma_{\br}^z+\sqrt{3}\sigma_{\br}^x),\qquad
\hat\pi_{\br}^y=\tfrac14(-\sigma_{\br}^z-\sqrt{3}\sigma_{\br}^x)
\end{equation}
and
\begin{equation}
\hat\pi_{\br}^z= \tfrac12\sigma_{\br}^z.
\end{equation}
which defines the \emph{120$^\circ$-model} on the level of a quantum spin system. 

In the $\tg$ compounds (e.g., LaTiO$_3$) the general form of \twoeqref{1}{orbHam} is preserved but the appropriate choice of the $\hat\pi_{\br}^\alpha$'s is now $\hat\pi_{\br}^\alpha = \frac12 \sigma_{\br}^\alpha$ for $\alpha=x,y,z$, see~\cite{Khal}. This is called the \emph{orbital compass} model. It is worth noting that an accounting of the strain field in the $\eg$ compounds leads directly to orbital interactions of the 120$^\circ$-type, see~\cite{Khomskii}, while if the strain fields are introduced in the~$\tg$ cases, the upshot is yet another orbital-only term akin to those discussed so far. Notwithstanding, in the~$\tg$ cases our analysis is largely incomplete and so we will confine the bulk of our attention to the 120$^\circ$-model. 

Throughout this Letter we will only discuss the \emph{orbital-only} models in which the spin degrees of freedom are suppressed. This approach may be presumed to capture the essential orbital physics of the systems at hand, cf~Refs.~\cite{brink03,kubo}. We remark that in all of these models, ordering among the spins is not necessarily a question of 
pertinence. In particular, in the itinerant-electron version of the orbital compass model, the elegant 
Mermin-Wagner argument of Ref.~\cite{Harris03} apparently precludes this
possibility. However, the results in Ref.~\cite{Harris03} do \emph{not} 
preclude the physically relevant possibility of orbital ordering which, as we show in this Letter, is realized at least in the classical versions of these~systems.

Henceforth, we will deal only with the orbital pseudo-spins 
which we denote by~$\bS_{\br}$ instead of $\hat\pi_{\br}$. 
In the context of the orbital-only models, we consider the standard $S\to\infty$ finite temperature limit. 
As is well known~\cite{Lieb-Simon}, this results in the classical analogues 
of the respective Hamiltonians, where the quantum variables 
are replaced by classical two or three-component spins. 
We proceed with a concise definition.

\section{Classical orbital-only models}
We start with the 120$^\circ$-model which is the most prominent of all of the above. The model is defined on the usual cubic lattice where at each site~$\br$ there is a unit-length two-component 
spin (associated with the 
two dimensional $\eg$ subspace) denoted by~$\bS_{\br}$. 
Let $\hata$, $\hatb$ and~$\hatc$ denote three evenly-spaced vectors on the unit circle separated by~120 degrees. To be specific let us have~$\hata$ point at~$0^\circ$ with~$\hatb$ and~$\hatc$ pointing at~$\pm120^\circ$, respectively. We define the projection $S_{\br}^{(\hata)}=\bS_{\br}\cdot\hata$, and similarly for  $S_{\br}^{(\hatb)}$ and~$S_{\br}^{(\hatc)}$.
Then the 120$^{\circ}$ orbital model Hamiltonian is given by
\begin{equation}
\label{H120}
\mathscr{H}=-J\sum_{\br}\bigl(S_{\br}^{(\hata)}S_{\br+\hate_x}^{(\hata)}
+S_{\br}^{(\hatb)}S_{\br+\hate_y}^{(\hatb)}+S_{\br}^{(\hatc)}S_{\br+\hate_z}^{(\hatc)}\bigr),
\end{equation}
where the classical nature of the interaction always allows us to set $J>0$ 
\cite{ferro}.

The Hamiltonian of the orbital compass model has an 
identical form, i.e., we can still write
\begin{equation}
\label{Horig}
\mathscr{H}=-J\sum_{\br,\alpha}S_{\br}^{(\alpha)}S_{\br+\hate_\alpha}^{(\alpha)},
\end{equation}
only now~$\bS_{\br}$ are three-component spins 
and the superscripts represent the corresponding Cartesian components.
The seminal feature of both models is an infinite degeneracy of the ground state.
In particular, any constant spin-field, $\bS_{\br}\equiv\bS$, will be a ground state in both cases.
This is established by noting that 
$\sum_\alpha[S_{\br}^{(\alpha)}]^2$ is constant in both problems. 
Thus, up to an irrelevant constant, the general Hamiltonian of Eq.~\eqref{Horig}
is \begin{equation}
\label{Hdiff}
\mathscr{H}=\frac J2\sum_{\br,\alpha}\bigl(S_{\br}^{(\alpha)}-S_{\br+\hate_\alpha}^{(\alpha)}\bigr)^2,
\end{equation}
which is obviously minimized when~$\bS_{\br}$ is constant.
We emphasize that the continuous symmetries which underscore these ground states are just symmetries of the states and \emph{not} of the Hamiltonian itself. Therefore, at least in the classical orbital-only models, we are not in a setting where a Mermin-Wagner argument can be applied.

\newcounter{obrazek}

\begin{figure}[t]
\refstepcounter{obrazek}
\label{fig1}
\ifpdf
\centerline{\includegraphics[width=5.0in]{onetwenny.pdf}}
\else
\centerline{\includegraphics[width=5.0in]{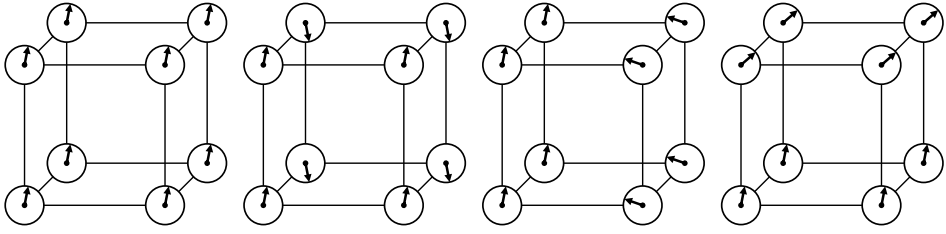}}
\fi
\medskip
\caption{The four possible ground states for the 120$^\circ$-model on a cube with one spin fixed.
The stratification 
structure of any (global) ground state is demonstrated by checking for consistency between all neighboring cubes.}
\end{figure}

Matters are further complicated because, as it turns out, the constant spin fields are \emph{not} the only ground states. Indeed, in the 120$^\circ$-model, starting from some constant-field ground state, another ground state may be obtained, e.g., by reflecting all spins in the~$xy$-plane through the vector~$\hatc$. This new state can be further mutated by introducing more flips of this type in other planes parallel to the~$xy$-plane. Obviously, similar alterations of the ``pristine'' states can take place in the other two coordinate directions. What is not so obvious, but nevertheless true~\cite{BCN2}, is that the abovementioned exhaust all the possible ground states for the 120$^\circ$-model: There is one direction of stratification (layering); the corresponding projection of~$\bS_{\br}$ \emph{is} constant throughout the system, leaving two possibilities for the other projections. In the various planes orthogonal to the stratification direction either of these choices can be independently implemented. This classification is proved by considering all possibilities of an elementary cube with a single spin fixed, and ensuring consistency in the tiling of the lattice; see Fig.~\ref{fig1}. The ground state situation for the orbital compass model is far more complicated and it will not be discussed till the end of this Letter.

\section{Spin-wave calculations}
Let us now investigate the effects of finite temperature.
Here, in general, we will see there is a fluctuation driven stabilization---sometimes known as ``order by disorder''~\cite{Shender&Henley}---that selects only a few of the ground states. The present arguments differ from the established standards, in part due to the complications caused by the stratified ground states.
We will focus on the 120$^\circ$-model. Here we can parameterize each spin~$\bS_{\br}$ by the angle~$\theta_{\br}$ with the $x$-axis.
In this language, let us consider the finite-temperature fluctuations about the ``pristine'' ground states where each $\theta_{\br}=\theta^\star$ (we will worry about the other ground states later).
At low temperatures, nearby spins will tend to be aligned, so we can work with the variables $\vartheta_{\br}=\theta_{\br}-\theta^\star$.
Neglecting terms of order higher than quadratic in~$\vartheta_{\br}$, the Hamiltonian \eqref{Hdiff} becomes
\begin{equation}
\mathscr{H}_{\SW}=\frac J2\sum_{\br,\alpha}q_\alpha(\theta^\star)(\vartheta_{\br}-\vartheta_{\br+\hate_\alpha})^2,
\end{equation}
where $\alpha=x,y,z$ while
$q_x(\theta^\star)=\sin^2(\theta^\star)$, $q_y(\theta^\star)=\sin^2(\theta^\star+120^\circ)$ and $q_z(\theta^\star)=\sin^2(\theta^\star-120^\circ)$.

Our preliminary goal is to compute the free energy as a function of~$\theta^\star$.
Let us assume that we are on a finite torus of linear dimension~$L$.
Interpreting~$\theta^\star$ as the \emph{average} of~$\theta_{\br}$ on the torus, we let $Z_L(\theta^\star)$ to denote the partition function
\begin{equation}
\label{part}
Z_L(\theta^\star)=\int
\delta\Bigl(\sum_{\br}\vartheta_{\br}=0\Bigr)\,e^{-\beta\mathscr{H}_{\SW}}
\prod_{\br}\frac{\textd\vartheta_{\br}}{\sqrt{2\pi}}.
\end{equation}
A standard Gaussian calculation then yields
\begin{equation}
\label{Gauss}
\log Z_L(\theta^\star)=-\frac12
\sum_{\bk\ne\boldsymbol0}\log\Bigl\{\sum_\alpha \beta J q_\alpha(\theta^\star)\,E_\alpha(\bk)\Bigr\},
\end{equation}
where $\bk=(k_x,k_y,k_z)$ is a vector in the reciprocal lattice and $E_\alpha(\bk)=2-2\cos k_\alpha$.
The right-hand side divided by~$L^3$ produces in the limit~$L\to\infty$ the (dimensionless) spin-wave free energy~$F(\theta^\star)$ for deviations around direction~$\theta^\star$. 
A tedious and rather unenlightening bit of analysis~\cite{BCN2} now shows that the spin-wave free energy~$F(\theta^\star)$ has \emph{strict} minima at $\theta^\star = 0^\circ$, $60^\circ$, $120^\circ$, $180^\circ$, $240^\circ$ and $300^\circ$. 

Let us now briefly discuss how the stratified states are handled. The key facts are as follows: (i)~A single interface between two types of ``pristine'' states generates an effective surface tension.
(ii)~The cumulative cost of many interfaces is effectively additive.
(iii)~The surface tension may be bounded by by the bulk free energy difference between the ``pristine'' state and period-two states; i.e., one in which there are as many interfaces as possible.
A full mathematical justification of all of the above would lead us too far astray---details are to be found in \cite{BCN2}---let us just compute the free energy of the period-two states. Specifically, let us consider the state which alternates between~$\theta\equiv\theta^\star$ and~$\theta\equiv-\theta^\star$ in the planes perpendicular to the~$x$ direction. In this case the limiting free energy is given by
\begin{equation}
\widetilde F(\theta^\star)=\frac14\int_{[-\pi,\pi]^3}
\frac{\textd\bk}{(2\pi)^3}\log\det\bigl(\,\beta J\Pi_{\bk}(\theta^\star)\bigr),
\end{equation}
where~$\Pi_{\bk}(\theta^\star)$ is the matrix
\begin{equation}
\Pi_{\bk}(\theta^\star)=
\left(\begin{aligned}
q_1 E_1+q_+E_+ & \qquad q_-E_-\\
q_-E_-\qquad & q_1 E_1^\star+q_+E_+
\end{aligned}\right).
\end{equation}
Here we have let~$q_\alpha=q_\alpha(\theta^\star)$ and~$E_\alpha=E_\alpha(\bk)$ be as above, and we have abbreviated
\begin{equation}
E_\alpha^\star(\bk)=E_\alpha(\bk+\pi\hate_\alpha),\quad
q_\pm=\frac12(q_2\pm q_3)\quad\text{and}\quad
E_\pm=E_2\pm E_3.
\end{equation} 
An elementary convexity analysis shows that~$\widetilde F(\theta^\star)>F(0^\circ)$ for~$\theta^\star\ne0^\circ,180^\circ$ (while, as is readily checked, $\widetilde F(0^\circ)$ equals~$F(0^\circ)$). 

Thus, at the level of spin-wave approximation, it is clear that finite-temperature effects will select six ground states above all others. Of course, this is only the beginning of a complete mathematical analysis: One must account for all other possible thermal disturbances and their interactions, the interactions of said additional disturbances with the spin waves and, not to mention, the interaction of spin waves with one another. Any such approach is, of course, hopeless even at the level of perturbation theory. Indeed, as can be readily verified, the latter is beset with infrared divergences even at the lowest non-vanishing order.

\section{Sketch of rigorous proof}
Our approach~\cite{BCN2}, which automatically circumvents these (and other unnamed) difficulties, proceeds as follows: First, we partition the lattice into blocks of side~$B$, generically denoted by~$\Lambda_B$, where~$B$ is a scale to be determined later. Then we pick a small number~$\kappa>0$ and call a block~$\Lambda_B$ \emph{good} if the spin configuration on the block is everywhere within~$\kappa$ of one of six ground states mentioned above. We will use~$\GG$ to denote the event that the block is good. Clearly, for~$\kappa\ll1$, there are six disjoint good-block events~$\GG_0,\GG_{60},\dots,\GG_{300}$. Blocks that are not good will be referred to as \emph{bad} and the corresponding event will be denoted by~$\BB$.

The goal of our analysis is to show that (i) most blocks are good and (ii) it is unlikely that any given pair of good blocks---no matter their separation---are of distinct types of goodness. For~(i) it suffices to show that the event~$\BB$ has very small probability. Here we introduce yet another scale~$\Delta$ (with~$\Delta\ll\kappa$) and decompose~$\BB$ according to the pertinent reason for the badness of the corresponding block. Specifically, we write~$\BB$ as the disjoint union
\begin{equation}
\BB=\BBE\cup\BBSW.
\end{equation}
Here~$\BBE$ marks the situation in which an ``energetic disaster'' has occurred inside the block, i.e., there is a nearest-neighbor pair of spins such that $|S_{\br}^{(\alpha)}-S_{\br+\hate_\alpha}^{(\alpha)}|>\Delta$, while~$\BBSW$ is the event that the energetics is good---which implies that the configuration in the block is near \emph{some} ground state---but the spin-wave entropy is not as good as for the ``good'' ground states. To estimate the probability of~$\BB$ we now show that~$\BbbP_\beta(\BBE)$ is suppressed like~$B^3e^{-\beta\Delta^2}$ while~$\BbbP_\beta(\BBSW)$ is suppressed exponentially in powers of~$B$. (We will get to the details of these estimates momentarily.) Both of these are small if~$B$ is large and~$\beta\Delta^2\gg\log B$.

The aforementioned arguments establish that bad blocks are unlikely to appear; but a variant of these estimates also proves that the type of goodness will be uniform throughout the system. Indeed, suppose two blocks are of distinct type of goodness. A moment's thought now shows that to get from one block to the other along any path we must either pass through an energetically charged block or visit a block with non-ideal spin-wave entropy. We conclude that the two good blocks are separated by a ``barrier'' of bad blocks. As it turns out, the probability of any given barrier is bounded by the product of probabilities to get the constituting blocks. Thus, a barrier is suppressed exponentially in its size and, ultimately, we can appeal to a standard Peierls' argument to finish the proof. 

Having sketched the backbone of our argument, let us pause to explain how the bad-block estimates are technically implemented. Here we call upon the standard technique of \emph{chessboard estimates}~\cite{FSS-FILS}. These go roughly as follows: To each (reflection-symmetric) event~$\AA$ which can take place in the block~$\Lambda_B$ we may define the quantity~$\zz_\beta(\AA)$ which is the partition function per site computed under the constraint that~$\AA$ occurs in every translate of~$\Lambda_B$ by integer multiples of~$B$. Then the thermal-state probability of observing~$\AA$~is bounded by
\begin{equation}
\label{CE}
\BbbP_\beta(\AA)\le\Bigl(\frac{\zz_\beta(\AA)}{\zz_\beta}\Bigr)^{B^3}
\end{equation}
where~$\zz_\beta$ is the unconstrained partition function per site. Moreover, the probability of the simultaneous occurrence of $n\ge1$ translates of the event~$\AA$ is bounded by the right-hand side of~Eq.~\eqref{CE} \emph{raised to the $n$-th power}. 

An application of the above technology to the event~$\AA=\BBE$ directly yields the bound $\BbbP_\beta(\BBE)\le c_1B^de^{-\beta\Delta^2}$ where~$c_1$ is a constant. As for the event~$\BBSW$, if the corresponding block is ``pristine'' then, in the computation of~$\zz_\beta(\AA)$, we may assume that the harmonic approximation is ``good'' and the spin-wave calculations from the previous sections (up to small errors) may be applied. Thus, terms of this form are suppressed like~$e^{-c_2B^3}$. If the bad block has interfaces, more refined chessboard-type arguments are employed and the upshot is a suppression of the form~$e^{-c_3B^2}$. Combining these bounds, the desired estimates are readily proved.

The situation in the orbital-compass model is considerably more complicated due to the profusion of additional ground states. Here, starting from a homogeneous ground state, the spins in an entire plane can be continuously rotated about the axis perpendicular to that plane without any disruption of the energetics. Thus, unlike in the 120$^\circ$-model, an elementary cube with one spin fixed has a \emph{continuum} of distinct ground states. Notwithstanding, it has been established~\cite{BCN1} that even in this case orbital ordering occurs. However, the nature of the thermal states differs, in certain details, from that of the 120$^\circ$-model. For instance,~$\langle\bS_{\br}\rangle$ vanishes at each site with the ordering being something along the lines of a nematic type.

\section{Conclusion}
We have demonstrated that the classical 120$^\circ$-model exhibits long-range order at sufficiently low temperatures. The key feature is that the degeneracy of the ground states is broken at positive temperatures via a kind of ``order-by-disorder'' mechanism. A complete argument, on a level of mathematical theorems, has already been constructed for the~120$^\circ$-model~\cite{BCN2} and similar (albeit less explicit) results hold for the orbital compass model~\cite{BCN1}.  All of this strongly indicates that there is orbital ordering in the full-blown quantum/itinerant-electron versions of these orbital models, wherein zero point fluctuations might further stabilize this orbital order. 

\smallskip\noindent
\textit{Acknowledgments.}~This research was supported by NSF DMS-0306167
(M.B. \& L.C.), US DOE via LDRD X1WX (Z.N.) and FOM~(J.B.).


\begin{thebibliography}{aaa}
\bibitem{ferro} 
Reflection of every even spin shows that the classical~$J>0$ and~$J<0$ 120$^\circ$-models are equivalent. Due to the absence of any~$\sigma_y$-terms, a similar transformation works even for the quantum version of the 120$^\circ$-model.
\end{thebibliography}
\end{document}